\documentclass[prd,twocolumn] {revtex4}
\usepackage{graphicx}
\usepackage{epsfig}
\usepackage{bm}
\usepackage{amssymb}
\usepackage{float}
\usepackage{amsmath}
\usepackage{dcolumn}
\usepackage{cancel}
\usepackage{color}
\usepackage{epstopdf}
\usepackage{nccbbb}
\usepackage{lipsum}
\UseRawInputEncoding

\begin{document}

\title {Can interacting dark energy with dynamical coupling resolve the Hubble tension}

\author{Yan-Hong Yao}
\email{yhy@mail.nankai.edu.cn}

\affiliation{School of Physics and Astronomy, Sun Yat-sen University,
2 Daxue Road, Tangjia, Zhuhai, People¡¯s Republic of China}

\author{Xin-He Meng}
\email{xhm@nankai.edu.cn}

\affiliation{School of Physics, Nankai University, Tianjin, 300071, People¡¯s Republic of China}

\begin{abstract}
The $H_0$ tension between low- and high- redshift measurements is definitely a serious issue faced by current cosmologists since it ranges from 4$\sigma$ to 6$\sigma$. To relieve this tension, in this paper we propose a new interacting dark energy model with time varying coupling parameter by parameterizing the densities of dark matter and dark energy, this parametric approach for interacting dark sectors are inspired by our previous work concerning the coupled generalized three-form dark energy model in which dark matter and dark energy behave like two uncoupled dark sectors with effective equation of state when the three-form $|\kappa X|\gg1$, for this reason, we reconstruct coupled generalized three-form dark energy from such parametric model under the condition $|\kappa X_0|\gg1$. In the end, we place constraints on three parametric models in light of the Planck 2018 cosmic microwave background (CMB) distance priors, baryon acoustic oscillations (BAO) data and the Pantheon compilation of Type Ia supernovae (SN Ia) data by assuming the parameter $k$ as 0,5,10 respectively. The fitting results show that, for all the observational data sets, the parametric models with $k=0,5,10$  relieve the Hubble tension with the latest local determinations of the Hubble constant from ${\rm SH_0ES}$ team, i.e. the so called R22, to 2.3 $\sigma$, 1.9 $\sigma$ and 1.3 $\sigma$ with $\chi_{\rm min}^2=10.43,10.47,10.48$ respectively, revealing that $k$ is positive correlated to the Hubble constant.

\textbf{}
\end{abstract}

\maketitle

\section{Introduction}
\label{intro}
\cite{aghanim2020planck,asgari2021kids,amon2022dark,secco2022dark,philcox2022boss,li2019simple,agrawal2019rock,yao2018power}Despite the $\Lambda$ - Cold Dark Matter ($\Lambda$CDM) model has proven to provide an excellent fit to a wide range of cosmological data\cite{riess1998observational,perlmutter1999measurements,dunkley2011atacama,hinshaw2013nine,story2015measurement,Beutler2011The,ross2015clustering,alam2017clustering,troxel2018dark}, many open questions remain, including the very nature of dark matter and dark energy, which both play an important role to the evolution of the current universe. Furthermore, as astronomers improve the precision of cosmological observations, tensions among the values of cosmological parameters inferred from different datasets have also emerged.

The most striking discrepancy is between the values of Hubble constant $H_0$ inferred from Planck Cosmic Microwave Background (CMB) data assuming the $\Lambda$CDM model, $H_0=67.4\pm0.5$ ${\rm km\cdot s^{-1}Mpc^{-1}}$ \cite{aghanim2018planck}, and the latest distance measurements of supernovaes calibrated by Cepheid variables, $H_0=73.04\pm1.04$ ${\rm km\cdot s^{-1}Mpc^{-1}}$ \cite{riess2022comprehensive}(denoted as R22 hereafter), these two values are differ by $\simeq$ 5$\sigma$ from each other. In addition to the Hubble tension, other anomalies also emerge, including the tension in $S_8=\sigma_8\sqrt{\Omega_{\rm m}/0.3}$ (where $\Omega_{\rm m}$ is the matter density parameter and $\sigma_8$ is the matter fluctuation amplitude on scales of ${\rm 8h^{-1}Mpc}$) between low-redshift probes such as weak gravitational lensing and galaxy clustering \cite{asgari2020kids+,asgari2021kids,joudaki2020kids+,abbott2022dark,amon2022dark,secco2022dark,loureiro2021kids,joudaki2017kids,heymans2021kids,hildebrandt2020kids+,abbott2020dark,macaulay2013lower,skara2020tension,joudaki2016cfhtlens,bull2016beyond,kazantzidis2018evolution,nesseris2017tension,philcox2022boss} and Planck CMB measurements, which ranges from 2$\sigma$ to 3$\sigma$, and the preference of the Planck CMB angular spectra for a large amplitude of the lensing signal $A_{\rm lens}$, with $A_{\rm lens}>1$ more than 95 \% C.L. \cite{aghanim2018planck} ($A_{\rm lens}$ equals to 1 in the $\Lambda$CDM model). Some of these discrepancies may due to considerable unaccounted systematic errors in the Planck observation and/or the local distance ladder measurements, however, this is not the case for the Hubble tension, indeed, as shown in \cite{krishnan2020there,krishnan2021running,dainotti2021hubble,dainotti2022evolution,colgain2022revealing,colgain2022putting}, there is an evidence for a descending trend in the Hubble constant. Krishnan et al. in \cite{krishnan2020there} considered data set comprising megamasers, cosmic chronometers (CC), SN Ia and baryonic acoustic oscillations (BAO), and binned them according to their redshift. By fitting $H_0$ with each bin assuming the $\Lambda$CDM model, they showed that $H_0$ descends with redshift, allowing one to fit a line with a non-zero slope of statistical significance 2.1$\sigma$. In addition, Dainotti et al. in \cite{dainotti2021hubble,dainotti2022evolution} analyzed the Hubble tension in the Pantheon sample through a binning approach, and also obtained a slowly decreasing trend for $H_0$. The key point here is that there are only two possibility which can result in Hubble tension, i.e. systematic errors and new physics, if the Hubble tension is due to systematic errors, then one doesn't expect to see a running $H_0$. Therefore, the Hubble constant tension does call for new physics. In the following discussion, we focus on the Hubble tension, other anomalies are not within the scope of our discussion.

There are many different kinds of new models proposed to replace the $\Lambda$CDM model, almost all of these models can be divided into two categories, i.e. early-time solutions and later-time solutions, the former refer to the modifications
of the expansion history before recombination, and the latter, however, modify the expansion history after the recombination. Among all of the early-time solutions, the early dark energy model proposed in \cite{poulin2019early} is the most well-known one, such model introduces an exotic early dark energy played by a scalar field $\phi$ with a potential having an oscillating feature, it acts as a cosmological constant before a critical redshift $z_c (z_c\gtrsim3000)$ but whose density then dilutes faster than radiation, has
been shown to be an effective possibility for reducing the $H_0$ tension. Other early dark energy models such as those proposed in \cite{agrawal2019rock,smith2020oscillating,lin2019acoustic,niedermann2021new,freese2021chain,ye2020hubble,akarsu2020graduated,braglia2020unified} have also showed their ability to reduce the Hubble tension. Besides early dark energy models, there are some other early-time solutions, the most famous one is known as the dark radiation model, i.e. a model with extra relativistic degrees of freedom $N_{\rm eff}$ beyond 3.046, Although such model can help with the $H_0$ tension, it is worth to mention that it is disfavored from both BAO and SN Ia data from a model comparison point of view \cite{vagnozzi2020new}.
Late-time solutions, although they are not considered to be as effective as the early-time solutions to alleviate the Hubble tension\cite{mortsell2018does}, these models,  including $w$CDM \cite{aghanim2018planck}, $w_0w_a$CDM \cite{aghanim2018planck}, model with negative dark energy density\cite{visinelli2019revisiting}, and interacting dark energy model \cite{yao2021relieve,yao2020new1,kumar2016probing,kumar2017echo,kumar2019dark,nunes2022new,yang2018tale,di2017can,yang2018interacting,di2020nonminimal,di2020interacting,cheng2020testing,lucca2020tensions,gomez2020update,yang2019dark,yang2020dynamical}, are also well-studied.

In \cite{yao2021relieve}, we proposed a coupled generalized three-form dark energy model (a generalization of the three-form dark energy models proposed in \cite{yao2018a,yao2020coupled}), in which dark energy is represented by a generalized three-form field and dark matter is represented by point particles that can be regarded as dust. The analysis of this model using Planck 2018 CMB distance priors, BAO data, JLA sample gives $H_0=70.1_{-1.5}^{+1.4}{\rm km s^{-1}Mpc^{-1}}$, alleviating the Hubble tension with R19 at 2.0$\sigma$. Such model can be regarded as an interacting dark energy model with  constant coupling parameter when the three-form $|\kappa X_0|\gg1$, which is supported by observation considered. To test whether a time varying coupling parameter is helpful to relieve the Hubble tension, in this paper we propose a new interacting dark energy model, with new parameters $\alpha,\beta$ and $k$, by parameterizing the densities of dark matter and dark energy, after implementing constraints on three parametric models specified by setting parameter $k$ as 0,5,10 against Planck 2018 CMB distance priors, BAO measurements and Pantheon dataset, we will see that $k$ is positive correlated to the Hubble constant for the CMB+Pantheon data set and the full data set.

The rest of this paper is organized as follows. In section \ref{sec:1}, we present a new interacting dark energy model by parameterizing the densities of dark matter and dark energy and then deduce the background evolution behavior of this model. In section \ref{sec:2}, we reconstruct coupled generalized three-form dark energy from such parametric model under the condition $|\kappa X_0|\gg1$. In section \ref{sec:3}, we confront three models with Planck 2018 CMB data, BAO data as well as SN Ia data and assess their ability to relieve the $H_0$ tension. In the last section, we make a brief conclusion with this paper.
\section{A new interacting dark energy model}
\label{sec:1}
In this section we introduce a new interacting dark energy model by assuming the densities of two dark sectors have the following form:
\begin{eqnarray}
  \Omega_{\rm c}(z) &=& \Omega_{\rm c}(1+z)^{3(1+w_{\rm CDM})},\\
  \Omega_{\rm X}(z) &=& \Omega_{\rm X}(1+z)^{3(1+w_{\rm DE})},\\
   w_{\rm CDM} &=&\frac{\int_0^N w_{\rm cdm}dN}{N}=\beta\frac{(1+z)^2}{(1+z)^2+k^2},\\
   w_{\rm DE}& = &w_{\rm de}= -1-\alpha.
\end{eqnarray}
Where $w_{\rm cdm}$ and $w_{\rm de}$ are the effective equations of state for two dark sectors which describe the equivalent uncoupled model in the background, $w_{\rm CDM}$ and $w_{\rm DE}$ are the average of them with respect to the variable $N={\rm ln}a$, $a$ is the scale factor. $\Omega_{\rm c}$ and $\Omega_{\rm X}$ are the density parameters of two dark sectors, since we also consider the evolution history before matter dominated era, we have $\Omega_{\rm X}+\Omega_{\rm r}+\Omega_{\rm b}+\Omega_{\rm c}=1$, $\Omega_{\rm r}=\Omega_{\gamma}(1+\frac{7}{8}(\frac{4}{11})^{\frac{4}{3}}N_{\rm eff})(N_{\rm eff}=3.046, \Omega_{\gamma}=\frac{(T_{\rm CMB}/2.7 {\rm K})^{4}}{42000h^2}, T_{\rm CMB}=2.7255{\rm K}, h=\frac{H_0}{100\rm km\cdot s^{-1}Mpc{-1}})$ and $\Omega_{\rm b}$ are the density parameters of radiation and baryons. One notes that, for $(1+z)\ll k$, we have $w_{\rm CDM}\ll \beta$, considering $\beta\ll1$ in any reasonable case, the strength of coupling of two dark sectors are negligible under this condition, for $(1+z)\gg k$, we have $w_{\rm CDM}\approx\beta$, in this case the coupling can be strong if $\beta\gtrsim0.01$. Now we have the following Hubble parameter:
\begin{equation}
\begin{split}
 E^2=&\frac{H^2}{H_0^2}
 =\Omega_{\rm r}(1+z)^{4}+\Omega_{\rm c}(1+z)^{3(1+w_{\rm CDM})}+\Omega_{\rm b}(1+z)^{3}\\
 &+(1-\Omega_{\rm r}-\Omega_{\rm b}-\Omega_{\rm c})(1+z)^{3(1+w_{\rm de})}.
\end{split}
\end{equation}
In order to obtain the expression for the coupling parameter, we first derived the formula for $w_{\rm cdm}$ by taking the derivative of the $Nw_{\rm CDM}$ with respect to $N$, which leading to
\begin{equation}
  w_{\rm cdm} = \beta\frac{(1+z)^2((1+z)^2+k^2(1+2{\rm ln}(1+z)))}{((1+z)^2+k^2)^2},
\end{equation}
Then by applying the energy conservation equations for two dark sectors, we obtain
\begin{eqnarray}
  \dot{\rho}_{\rm c}+3H\rho_{\rm c} &=& \delta H\rho_{\rm c},\\
  \dot{\rho}_{\rm X}+3(1+w_{\rm X})H\rho_{\rm X}&=& -\delta H \rho_{\rm c},
\end{eqnarray}
where the dark energy equation of state $w_X$ and the coupling parameter $\delta$ have the following form:
\begin{eqnarray}
  \delta &=& -3\beta\frac{(1+z)^2((1+z)^2+k^2(1+2{\rm ln}(1+z)))}{((1+z)^2+k^2)^2},\hspace{0.5cm} \\
  w_{\rm X} &=&-1-\alpha-\frac{\delta}{3}\frac{\rho_{\rm c}}{\rho_{\rm X}},
\end{eqnarray}
one notes that we obtain decaying dark matter if $\beta>0$, unlike most of other interacting dark energy model, the dark energy equation of state $w_X$ is modified by the coupling. One thing that cosmologist generally know is that when $w_X$ and $\delta$ are constant, there are some instabilities in the dark sector perturbations at early times if $w_X>-1$ \cite{valiviita2008large,he2009stability,gavela2009dark}. To investigate the early-time stability of our interacting dark energy model, let's recall the doom factor ${\rm \textbf{d}}$ for interaction with coupling function proportional to dark matter density, which is defined as \cite{gavela2009dark}:
\begin{equation}\label{}
  {\rm \textbf{d}}\equiv\frac{\delta}{3(1+w_{\rm X})}\frac{\rho_{\rm c}}{\rho_{\rm X}},
\end{equation}
\cite{gavela2009dark} shows that the sign of ${\rm \textbf{d}}$ defines the (un)stable regimes, for ${\rm \textbf{d}}>0$, instabilities can develop at early times, for ${\rm \textbf{d}}<0$, no instabilities are expected. Substituting the formulas of $\delta$ and $w_X$ into equation (11), and assuming $z\gg1$, we obtain ${\rm \textbf{d}}\rightarrow-1$, therefore the parametric model we proposed in this section is free from early-time instability regardless of the size and sign of $\alpha$ and $\beta$.

In order to investigate the effective dark energy with dust dark matter which describe the equivalent uncoupled model in the background, let's rewrite the Hubble parameter as follows:
\begin{equation}
   E^2 =\Omega_{\rm r}(1+z)^{4}+\Omega_{\rm m}(1+z)^{3}+(1-\Omega_{\rm r}-\Omega_{\rm m})\frac{\rho_{\rm X,eff}}{\rho_{\rm X0,eff}},
\end{equation}
where
\begin{widetext}
\begin{equation}\label{}
\frac{\rho_{\rm X,eff}}{\rho_{\rm X0,eff}}=\frac{1}{1-\Omega_{\rm m}-\Omega_{\rm r}}(\Omega_{\rm c}((1+z)^{3(1+w_{\rm CDM})}-(1+z)^{3})
+(1-\Omega_{\rm r}-\Omega_{\rm m})(1+z)^{3(1+w_{\rm de})})
=\exp(\int_{0}^{z}\frac{3(1+w_{\rm X,eff}(\tilde{z}))}{1+\tilde{z}}d\tilde{z})
\end{equation}
\end{widetext}
is the normalized effective dark energy density, and
\begin{widetext}
\begin{equation}\label{}
w_{\rm X,eff}(z)=-1+\frac{\Omega_{\rm c}((1+w_{\rm cdm})(1+z)^{3(1+w_{\rm CDM})}-(1+z)^3)+(1-\Omega_{\rm m}-\Omega_{\rm r})(1+w_{\rm de})(1+z)^{3(1+w_{\rm de})}}{\Omega_{\rm c}((1+z)^{3(1+w_{\rm CDM})}-(1+z)^3)+(1-\Omega_{\rm m}-\Omega_{\rm r})(1+z)^{3(1+w_{\rm de})}}
\end{equation}
\end{widetext}
is the effective equation of state of dark energy assuming dust dark matter. An expression which at small redshifts tends to:
\begin{widetext}
\begin{equation}\label{}
  w_{\rm X,eff}(z)\approx-1-\alpha+\frac{\beta\Omega_{\rm c}}{(1+k^2)(1-\Omega_{\rm m}-\Omega_{\rm r})}+\frac{\beta\Omega_{\rm c}((1-\Omega_{\rm m}-\Omega_{\rm r})(6+10k^2+6\alpha(1+k^2)+3\beta)-3\beta\Omega_{\rm c})}{(1+k^2)^2(1-\Omega_{\rm m}-\Omega_{\rm r})^2}z,
\end{equation}
\end{widetext}
while $w_{\rm X,eff}\rightarrow\beta(\beta>0)$ and $w_{\rm X,eff}\rightarrow 0(\beta<0)$ at very large redshifts. The equation (15) shows that $w_{\rm X,eff}$ would have a peculiar divergent behavior for case with $\beta<0$ in which the denominator vanishes at redshift $z\sim O(1-10)$.

\section{Reconstruction of coupled generalized three-form dark energy from parametric interacting dark energy model}
\label{sec:2}
\begin{figure*}
	\centering
	\includegraphics[scale=0.55]{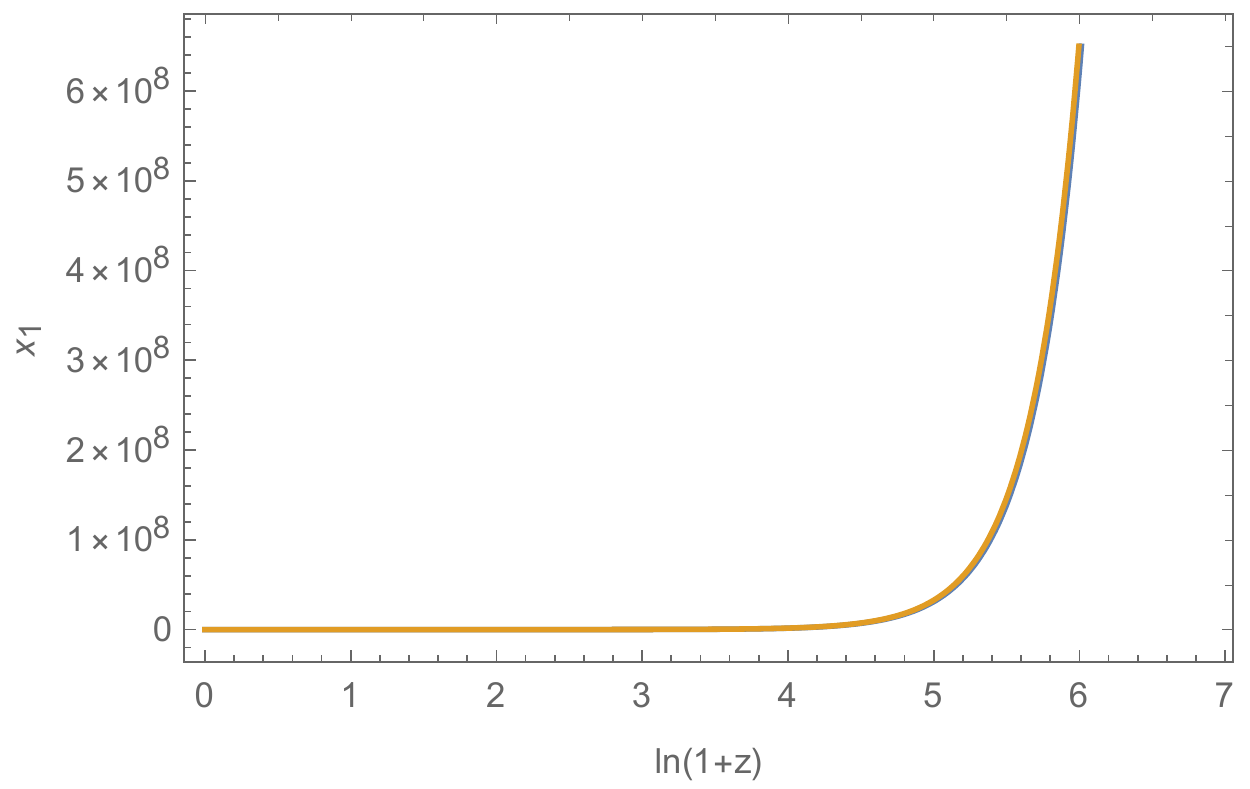}
	\includegraphics[scale=0.55]{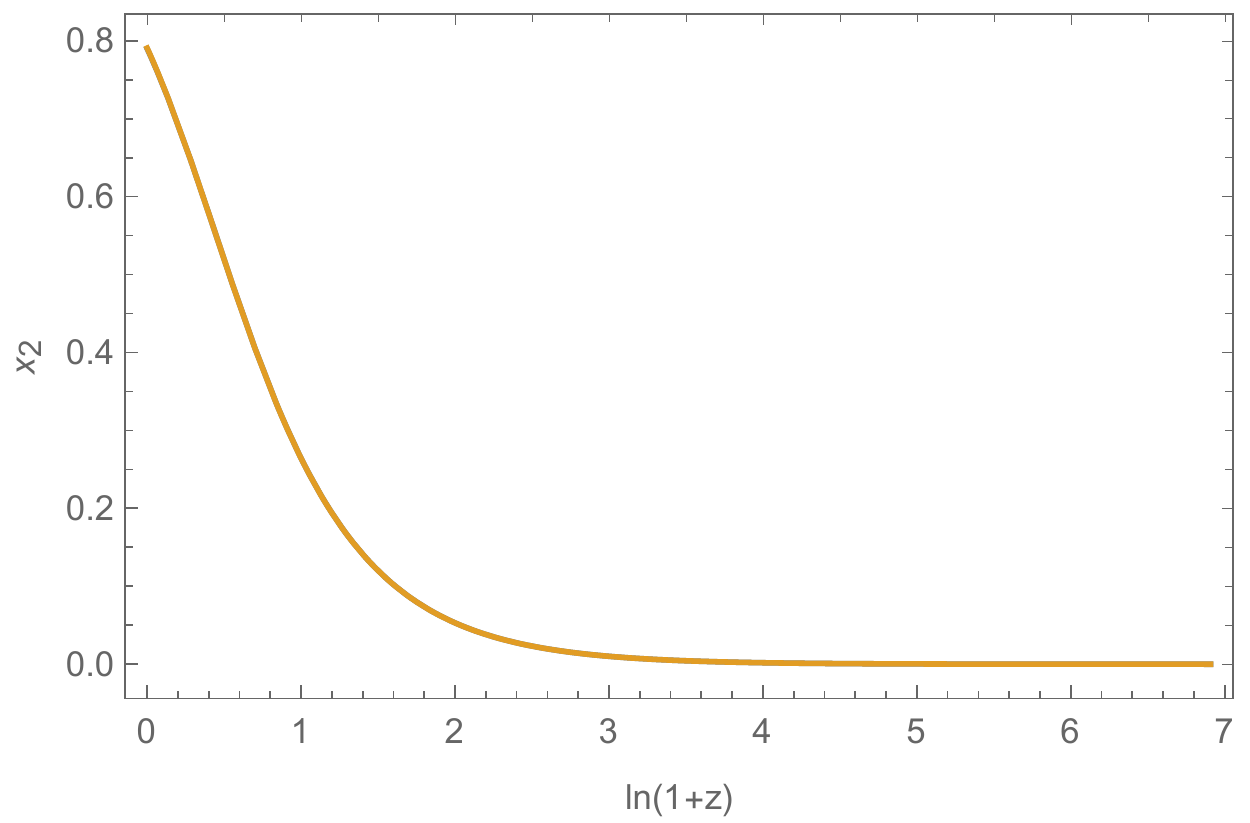}
	\caption{Evolution history of approximate and exact solutions of $x_1$ (left panel) and $x_2$ (right panel) with initial condition $x_{10}=10$ and $x_{20}=0.8$.}\label{fig:1}
\end{figure*}

Since in the coupled generalized three-form dark energy model proposed in our previous work, dark matter and dark energy behave like two uncoupled dark sectors with effective equation of state, which is similar to our parametric model assuming the three-form $|\kappa X|\gg1$, for this reason, we reconstruct coupled generalized three-form dark energy from such parametric model under the condition $|\kappa X_0|\gg1$ in this section. The total Lagrangian for a coupled generalized three-form dark energy model is written as
\begin{widetext}
\begin{equation}
\mathcal{L}=\frac{R}{2\kappa^2}-\frac{1}{48}F^{2}J(A^{2})-I(A^{2})\sum_{a}m_a\delta(x-x_a(t))\frac{\sqrt{-g_{\mu\nu}\dot{x}^\mu\dot{x}^\nu}}{\sqrt{-g}}+\mathcal{L}_{\rm b}+\mathcal{L}_{\rm r},
\end{equation}
\end{widetext}
where $R$ denotes the Ricci scalar and $\kappa=\sqrt{8\pi G}$ is the inverse of the reduced Planck mass. $A$ and $F = dA$ represent the three-form field and the field strength tensor, $J(A^2)$ and $I(A^2)$ contain total coupling information, which are the target functions we want to reconstruct, including the information of self-interaction of the three-form field and interaction between two dark sectors, respectively. $m_a$ denotes the mass of dark matter particles, b and r denote baryon and radiation, respectively. As usual, the universe is assumed as homogeneous, isotropic, and spatially flat, therefore it is described by the Friedmann-Robertson-Walker (FRW) metric:
\begin{equation}\label{}
  ds^{2}=-dt^{2}+a(t)^{2}d\vec{x}^{2},
\end{equation}
here $a(t)$ stands for the scale factor.
To be compatible with FRW symmetries, the three-form field is chosen as the time-like component of the dual vector field\cite{Koivisto2009Inflation2}, i.e.
\begin{equation}
  A_{i j k}=X(t)a(t)^{3}\varepsilon_{ijk},
\end{equation}
it has been normalized by the $a(t)^3$ so that $A^2 = 6X^2$. As a result, $J$ and $I$ can be regarded as functions of the field $X$. Now we define the shorthand notations for convenience
\begin{equation}\label{}
  g\equiv\frac{1}{\kappa}\frac{d{\rm ln}J}{dX}, f\equiv\frac{1}{\kappa}\frac{d{\rm ln}I}{dX},
\end{equation}
the Friedmann equations then can be written as
\begin{equation}\label{}
  H^{2} = \frac{\kappa^{2}}{3}(\frac{1}{2}J(\dot{X}+3HX)^{2}+\rho_{\rm c}+\rho_{\rm b}+\rho_{\rm r}),
\end{equation}
\begin{equation}\label{}
\begin{split}
\dot{H} &=-\frac{\kappa^{2}}{2}(-\frac{1}{2}\kappa X g J(\dot{X}+3HX)^{2}+(1+\kappa X f) \rho_{\rm c}\\&+\rho_{\rm b}+\frac{4}{3}\rho_{\rm r}),
\end{split}
\end{equation}

equation of motion for the three-form field is
\begin{equation}\label{}
\begin{split}
   &J(\ddot{X}+3\dot{H}X+3H\dot{X})+\frac{\kappa g J}{2}(\dot{X}-3HX)(\dot{X}+3HX)\\&+ \kappa f \rho_{\rm c}=0,
\end{split}
\end{equation}
and the continuity equation for two dark sectors is
\begin{eqnarray}
  \dot{\rho}_{\rm c}+3H\rho_{\rm c} &=&\kappa X^{\prime}fH \rho_{\rm c},\\
   \dot{\rho}_{\rm X}+3H(\rho_{\rm X}+p_{\rm X})& =& -\kappa X^{\prime}fH \rho_{\rm c},
\end{eqnarray}
where $\rho_{\rm X}$ and $p_{\rm X}$ are the density and pressure of the dark energy, which have the following form
\begin{eqnarray}
\rho_{\rm X} &=&\frac{1}{2}J(\dot{X}+3HX)^{2},\\
p_{\rm X} &=&-\frac{1}{2}J(1+\kappa X g  )(\dot{X}+3HX)^{2}+\kappa X f\rho_{\rm c},
\end{eqnarray}
from equation (25) and (26), we can derive the equation of state for dark energy as follows
\begin{equation}\label{}
  w_X=\frac{p_X}{\rho_X}=-1-\kappa X g +\kappa X f\frac{\rho_{\rm c}}{\rho_{\rm X}},
\end{equation}
which indicates that $w_X$ is modified by the coupling, just like the parametric model we introduced in last section. To discuss the background evolution behavior of the coupled generalized three-form dark energy model, it is convenient to introduce the following dimensionless variable
\begin{equation}\label{}
  x_{1}=\kappa X, x_{2}=\frac{\kappa}{\sqrt{6}}(X^{\prime}+3X), x_{3}=\frac{\kappa \sqrt{\rho}_{\rm b}}{\sqrt{3}H}, x_{4}=\frac{\kappa \sqrt{\rho_{\rm r}}}{\sqrt{3}H},
\end{equation}
and then it is easy to derive the following autonomous system from the background dynamical equations above
\begin{equation}
  x_{1}^{\prime}=\sqrt{6}x_{2}-3x_{1},
\end{equation}
\begin{equation}
  \begin{split}
  x_{2}^{\prime} & =(\frac{3}{2}(1+f x_1)(1-J x_{2}^2-x_{3}^2-x_{4}^2)-\frac{3}{2} g J x_1  x_{2}^2\\&+\frac{3}{2}x_{3}^2+2x_{4}^2)x_{2}
     -\frac{\sqrt{6}}{2 }gx_{2}^2+3g x_1x_{2}\\&
 -\frac{\sqrt{6}}{2} \frac{f}{J}(1-Jx_{2}^2-x_{3}^2-x_{4}^2),
\end{split}
\end{equation}
\begin{equation}
\begin{split}
  x_{3}^{\prime}&= -\frac{3}{2}x_{3}+(\frac{3}{2}(1+f x_1)(1-J x_{2}^2-x_{3}^2-x_{4}^2)\\&-\frac{3}{2} g J x_1  x_{2}^2+\frac{3}{2}x_{3}^2+2x_{4}^2)x_{3},
\end{split}
\end{equation}
\begin{equation}
\begin{split}
  x_{4}^{\prime}&= -2x_{4}+(\frac{3}{2}(1+f x_1)(1-J x_{2}^2-x_{3}^2-x_{4}^2)\\&-\frac{3}{2} g J x_1  x_{2}^2+\frac{3}{2}x_{3}^2+2x_{4}^2)x_{4},
\end{split}
\end{equation}
A prime here denotes derivation with respect to $N={\rm ln}a$. Now we have all the information associated with coupled generalized three-form dark energy model necessary for us to reconstruct functions $J$ and $I$ with respect to $x_1$, in order to do that, let's derive the following formula for Hubble parameter as a multivariate function of $z$, $(I/I_0)$ and $Jx_2^2$,
\begin{equation}
 E^2
 =\Omega_{\rm r}(1+z)^{4}+\Omega_{\rm c}(I/I_0)(1+z)^{3}+\Omega_{\rm b}(1+z)^{3}+J x_2^2E^2 ,
\end{equation}
where $I_0$ is the present value of $I$, comparing equation (33) with equation (5), we have the following corresponding expression for $(I/I_0)$ and $J$
\begin{equation}\label{}
  I/I_0=(1+z)^{3w_{\rm CDM}},
\end{equation}
\begin{equation}\label{}
   J= \frac{(1-\Omega_{\rm r}-\Omega_{\rm b}-\Omega_{\rm c})(1+z)^{3(1+w_{\rm de})}}{x_2^2E^2},
\end{equation}
Therefore, as long as we know the explicit expressions of $x_1$ and $x_2$ with respect to redshift $z$ by solving the autonomous system, we can reconstruct functions $J$ and $I$ with respect to $x_1$. In order to do that, we replace $f$, $g$, $(\frac{3}{2}(1+f x_1)(1-J x_{2}^2-x_{3}^2-x_{4}^2)-\frac{3}{2} g J x_1  x_{2}^2+\frac{3}{2}x_{3}^2+2x_{4}^2)$ and $1-Jx_{2}^2-x_{3}^2-x_{4}^2$ by the following corresponding expression.
\begin{equation}\label{}
  f=\frac{d{\rm ln}I}{dx_1}=\frac{d{\rm ln}(I/I_0)}{dx_1}=\frac{({\rm ln}(I/I_0))^{\prime}}{x_1^{\prime}}=\frac{-3w_{\rm cdm}}{\sqrt{6}x_{2}-3x_{1}},
\end{equation}
\begin{equation}\label{}
\begin{split}
g&= \frac{d{\rm ln}J}{dx_1}=\frac{({\rm ln}J)^{\prime}}{x_1^{\prime}}=\frac{-3(1+w_{\rm de})}{\sqrt{6}x_{2}-3x_{1}}+\frac{-2x_2^{\prime}}{x_2(\sqrt{6}x_{2}-3x_{1})}\\&-\frac{2E^{\prime}}{(\sqrt{6}x_{2}-3x_{1})E},
\end{split}
\end{equation}
\begin{equation}\label{}
\begin{split}
  &(\frac{3}{2}(1+f x_1)(1-J x_{2}^2-x_{3}^2-x_{4}^2)-\frac{3}{2} g J x_1  x_{2}^2+\frac{3}{2}x_{3}^2\\&+2x_{4}^2)=-\frac{E^{\prime}}{E},
\end{split}
\end{equation}
\begin{equation}\label{}
  1-Jx_{2}^2-x_{3}^2-x_{4}^2
     =\frac{\Omega_c(1+z)^{3(1+w_{\rm CDM})}}{E^2},
\end{equation}
with the help of these formulas, the autonomous system is reduced to the following two equations
\begin{equation}
  x_{1}^{\prime}=\sqrt{6}x_{2}-3x_{1},
\end{equation}
\begin{equation}\label{}
\begin{split}
  x_2^{\prime} =& (-1+\frac{\sqrt{6}x_2}{3x_1})(-\frac{E^{\prime}}{E}x_2-\frac{3x_1-\frac{\sqrt{6}}{2}x_2}{3x_1-\sqrt{6}x_2}(3\alpha-2\frac{E^{\prime}}{E})x_2\\
    & -\frac{3\sqrt{6}}{2}\frac{w_{\rm cdm}\Omega_c(1+z)^{3(1+w_{\rm CDM})}}{(3x_1-\sqrt{6}x_2)JE^2}),
\end{split}
\end{equation}
however, it is still impossible to find a analytical solution with the initial condition $x_1(0)=x_{10}$, $x_2(0)=x_{20}$ from this complicated system of equations. One technique that can be used to get rid of this difficult situation is to find approximate solution instead of exact solution from this system of equations. Therefore, we assume $|x_{10}|\gg1$ as well as $J>1$ under the condition $|x_{1}|\gg1$, surprisingly, we find that two equations above can be simplified to the following form
\begin{eqnarray}
   x_{1}^{\prime}&\approx&-3x_{1},\\
   x_2^{\prime}&\approx&(3\alpha+\frac{E^{\prime}}{E})x_2,
\end{eqnarray}
by solving these two equations, we obtain the following solution
\begin{eqnarray}
  x_1 &\approx&x_{10}(1+z)^3,\\
  x_2 &\approx& \frac{x_{20}(1+z)^{-3\alpha}}{E}.
\end{eqnarray}
To test the difference between approximate and exact solutions, we plot Fig.\ref{fig:1} by assuming $x_{10}=10$ and $x_{20}=0.8$, it can be inferred from Fig.\ref{fig:1} that the approximate and exact solutions basically coincide with each other, which shows that approximate solutions can give predictions that are close to exact solutions within a negligible error margin.

\begin{figure*}
	\centering
	\includegraphics[scale=0.55]{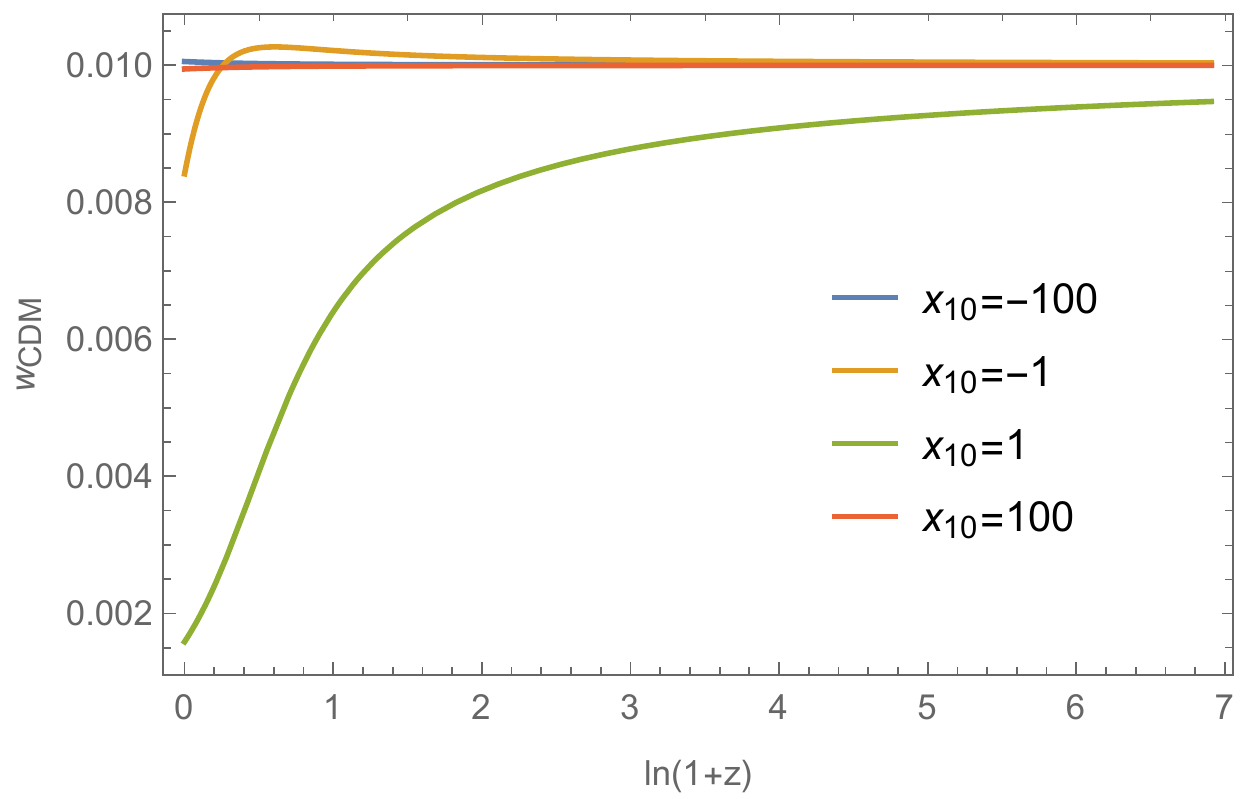}
	\includegraphics[scale=0.55]{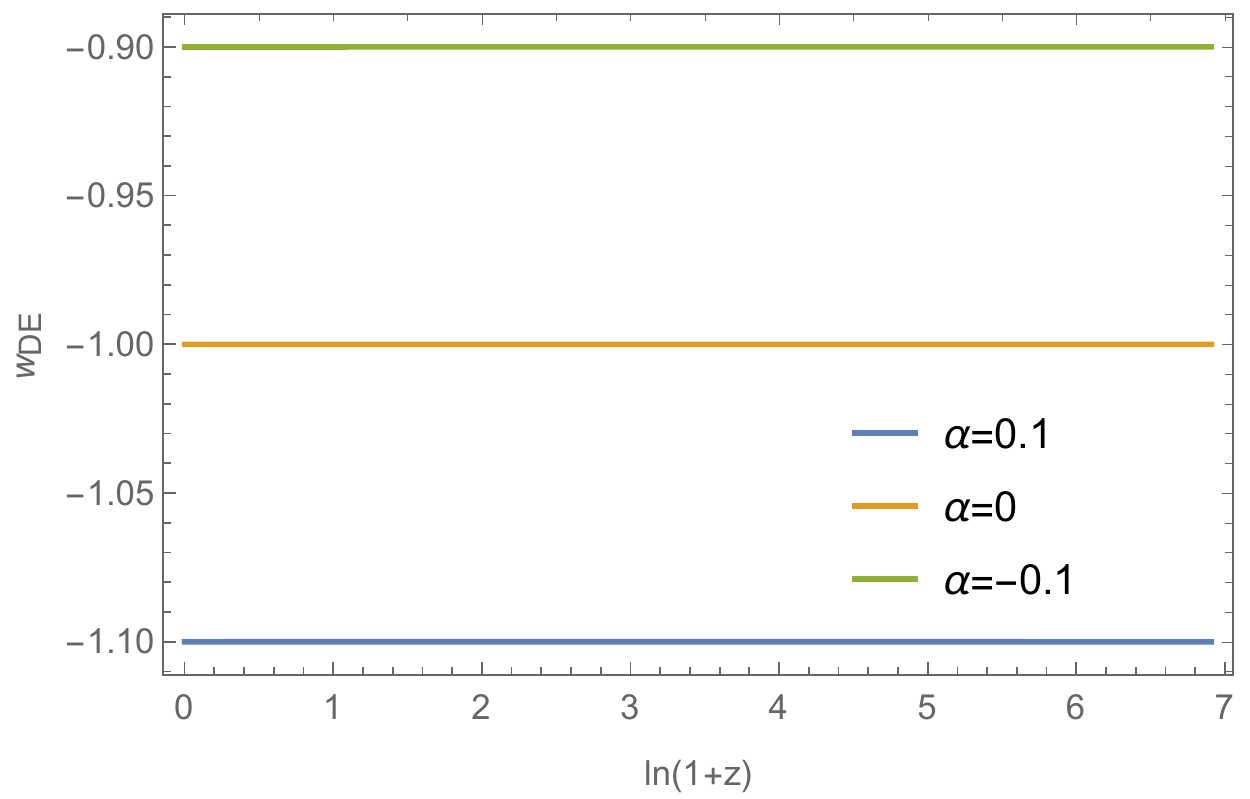}
	\caption{Evolution history of $w_{\rm CDM}$ (left panel) and $w_{\rm DE}$ (right panel) with respect to redshift ${\rm ln}(1+z)$. }\label{fig:2}
\end{figure*}

By substituting $x_2$ with its approximate solution in equation (35), we obtain an expression for $J$ with respect to $z$,
\begin{equation}\label{}
  J = \frac{(1-\Omega_{\rm r}-\Omega_{\rm b}-\Omega_{\rm c})(1+z)^{3\alpha}}{x_{20}^2}
\end{equation}
Now, with the help of the approximate solution of $x_1$, we finally can reconstruct $J(A^2)$ and $I(A^2)$ from the parametric model under the condition $|\kappa X_0|\gg1$
\begin{equation}\label{}
\begin{split}
I =  & I_0(\frac{x_1^2}{x_{10}^2})^{\frac{1}{2}\frac{\beta(x_1^2)^{\frac{1}{3}}}{(x_1^2)^{\frac{1}{3}}+k^{2}(x_{10}^2)^{\frac{1}{3}}}}=
I_0(\frac{\kappa^2A^{2}}{6x_{10}^2})^{\frac{1}{2}\frac{\beta(\kappa^2A^{2})^{\frac{1}{3}}}{(\kappa^2A^{2})^{\frac{1}{3}}+k^{2}(6 x_{10}^2)^{\frac{1}{3}}}} \\
=&I_0\exp({\frac{1}{2}\frac{\beta(\kappa^2A^{2})^{\frac{1}{3}}}{(\kappa^2A^{2})^{\frac{1}{3}}+k^{2}(6 x_{10}^2)^{\frac{1}{3}}}}{\rm ln}(\frac{\kappa^2A^{2}}{6x_{10}^2})),
\end{split}
\end{equation}
\begin{equation}\label{}
\begin{split}
    J&=\frac{(1-\Omega_{\rm r}-\Omega_{\rm b}-\Omega_{\rm c})(x_1^2)^{\frac{\alpha}{2}}}{(x_{10}^2)^{\frac{\alpha}{2}}x_{20}^2}\\&=\frac{(1-\Omega_{\rm r}-\Omega_{\rm b}-\Omega_{\rm c})(\kappa^2A^{2})^{\frac{\alpha}{2}}}{(6x_{10}^2)^{\frac{\alpha}{2}}x_{20}^2}\\&=J_0\frac{(\kappa^2A^{2})^{\frac{\alpha}{2}}}{(6x_{10}^2)^{\frac{\alpha}{2}}}.
\end{split}
\end{equation}
Where $J_0=\frac{(1-\Omega_{\rm r}-\Omega_{\rm b}-\Omega_{\rm c})}{x_{20}^2}$ is the present value of $J$. Therefore, for any taking values of $I_0$ and $J_0$ accompanied by a choice of a large value of $\sqrt{\kappa^2A_0^2/6}=|x_{10}|$, the generalized coupled three form dark energy model specify by (47) and (48) mimics the phenomenologically parameterized model proposed above with same values of parameters $\alpha$, $\beta$ and $k$. On the other hand, if we have explicit expressions of $J(A^2)$ and $I(A^2)$, we can determine the background evolution history by calculating the following two quantities
\begin{eqnarray}
  w_{\rm CDM} &=&\frac{{\rm ln}(I/I_0)}{3{\rm ln}(1+z)}, \\
  w_{\rm DE} &=& \frac{{\rm ln}(\frac{J x_2^2 E^2}{(1-\Omega_{\rm r}-\Omega_{\rm b}-\Omega_{\rm c})(1+z)^{3}})}{3{\rm ln}(1+z)},
\end{eqnarray}
for example, by assuming $J(A^2)$ and $I(A^2)$ as the following forms which specify the coupled generalized three-form dark energy model proposed in \cite{yao2021relieve}, we can obtain the evolution history of $w_{\rm CDM}$ and $w_{\rm DM}$ for such model.
\begin{eqnarray}\label{}
  J &=&(\alpha^2+\frac{\kappa^{2}}{6}A^{2})^{\frac{\alpha}{2}}=(\alpha^2+\kappa^{2}X^{2})^{\frac{\alpha}{2}} \hspace{1cm} \\
  I &=&(1+\frac{\kappa^{2}}{6}A^{2})^{\frac{\beta}{2}}=(1+\kappa^{2}X^{2})^{\frac{\beta}{2}}
\end{eqnarray}
In Fig.\ref{fig:2}, we plot $w_{\rm CDM}$  with respect to redshift ${\rm ln}(1+z)$ with the initial condition $\Omega_{\rm r}=0$, $\Omega_{\rm c}=0.26$, $\Omega_{\rm b}=0.04$, $\alpha=0.1$, $\beta=0.01$, $x_{10}=-100,-1,1,100$, and $w_{\rm DE}$ also with respect to redshift ${\rm ln}(1+z)$ with the initial condition $\Omega_{\rm r}=0$, $\Omega_{\rm c}=0.26$, $\Omega_{\rm b}=0.04$, $\alpha=0.1,0,-0.1$, $\beta=0.01$, $x_{10}=100$. It can be inferred from these two panels that with the choice of these initial condition we have $w_{\rm CDM}\approx\beta$ (leading to a constant coupling parameter as we already mentioned in the introduction) and $w_{\rm DE}\approx-1-\alpha$ when $|x_{10}|\gg1$. Since observations always prefers a large $|x_{10}|$, the parametric model proposed in last section can be regarded as the generalization of such coupled generalized three-form dark energy model.

\section{Confront the model with observations}
\label{sec:3}
In this section, instead of implementing constraints on the parametric interacting dark energy model proposed in section \ref{sec:1} (with newly introduced parameters $\alpha,\beta,k$), we perform a likelihoods analysis on three models, obtained from the parametric interacting dark energy model by setting $k=0,5,10$, using CMB distance priors, BAO data and SN Ia data since, for the parametric model with free $k$, the large error bar in $H_0$ can lead to an erroneous conclusion that the Hubble tension has been completely resolved considering $k$ is not only poorly constrained by the data but also correlated with $H_0$.

\subsection{CMB measurements}
For CMB observations, we use the Planck 2018 data release. In principle, one should apply the original Planck 2018 CMB temperature and polarization data directly to constrain the parameters in the models. However, thanks to the fact that the Hubble constant is only sensitive to the distance information extracted from CMB data, one can also use the distance priors from CMB data instead \cite{chen2019distance}. In this analysis, we would like to use the distance prior method for the purpose of saving time in programming and computation.
The distance priors is consist of two parameters, the shift parameter $R$ and the acoustic scale $l_{\rm A}$, the former reads
\begin{equation}
  R=\sqrt{\Omega_{\rm m}H_0^2}D_{\rm M}(z_{\ast}),
\end{equation}
where $D_{\rm M}(z_{\ast})$ is the comoving angular distance at decoupling, which depends on the dominant components after decoupling and is defined by
\begin{equation}\label{}
  D_{\rm M}=(1+z_{\ast})D_{\rm A}=\int_{0}^{z_{\ast}}\frac{dz}{H},
\end{equation}
the latter reads
\begin{equation}
  l_{\rm A}=\pi\frac{D_{\rm M}(z_{\ast})}{r_{\rm s}(z_{\ast})},
\end{equation}
in which $r_{\rm s}$ is the sound horizon at decoupling, which depends on the dominant components before decoupling and have the following form
\begin{equation}\label{}
  r_{\rm s}=\int_{0}^{t_{\ast}}\frac{c_{\rm s} dt}{a(t)}=\int_{0}^{a_{\ast}}\frac{c_{\rm s}da}{a^{2}H(a)},
\end{equation}
where $c_{\rm s}$ is the sound speed of fluid consisting of photons and baryons and is given by \cite{Efstathiou2010Cosmic}
\begin{equation}\label{}
  c_{\rm s}=\frac{1}{\sqrt{3(1+\frac{3\omega_{\rm b}a}{4\omega_{\gamma}})}}
\end{equation}
in which $\omega_{\rm b}=\Omega_{\rm b}h^2$, $\omega_{\gamma}=\Omega_{\gamma}h^2$. The redshift at decoupling $z_{\ast}$ can be express as \cite{Hu1996Small}
\begin{eqnarray}
  z_{*}&=&1048(1+0.00124\omega_{\rm b}^{-0.738})(1+g_{1}\omega_{\rm m}^{g_{2}}),\\
g_{1}&=&\frac{0.0783\omega_{\rm b}^{-0.238}}{1+39.5\omega_{\rm b}^{0.763}},\\
  g_{2}&=&\frac{0.56}{1+21.1\omega_{\rm b}^{1.81}},
\end{eqnarray}
in which $\omega_{\rm m}=\Omega_{\rm m}h^2$. These two parameters combined with baryon density $\omega_{\rm b}$ can provide a efficient extraction from full CMB data for us to place constraints on cosmological models.
The likelihood for the distance prior method then reads
\begin{equation}
 \chi_{\rm CMB}^{2}=s_{\rm CMB}^{T} C_{\rm CMB}^{-1} s_{\rm CMB},
\end{equation}
\begin{equation}
   s_{\rm CMB} = (R-1.7502,l_{\rm A}-301.471,\omega_{\rm b}-0.02236),
\end{equation}
where $C_{ij}=D_{ij}\sigma_i\sigma_j$ is the covariance matrix, $\sigma_{\rm CMB}=(0.0046,0.09,0.00015)$ is the errors, and
$D_{\rm CMB}=  \left(
      \begin{array}{ccc}
        1 & 0.46 &-0.66 \\
       0.46 & 1 & -0.33 \\
       -0.66&-0.33& 1\\
      \end{array}
    \right)$ is the covariance.

\subsection{Baryon acoustic oscillations}
For BAO observations, we employ the distance measurements from 6dFGS\cite{Beutler2011The}, SDSS-MGS\cite{ross2015clustering} and the BOSS DR12 \cite{alam2017clustering},  the total BAO likelihood reads
\begin{equation}
 \chi_{\rm BAO}^{2}=s_{\rm BAO}^{T} C_{\rm BAO}^{-1} s_{\rm BAO},
\end{equation}
with
\begin{widetext}
$C_{\rm BAO}=  \left(
      \begin{array}{cccccccc}
        0.00023 & 0 & 0 & 0 & 0& 0&0&0\\
       0 & 0.00007 & 0& 0 & 0& 0&0&0 \\
       0 &   0     & 624.7&23.7&325.3&8.3&157.4&3.6\\
       0&  0       &23.7& 5.6& 11.6& 2.3& 6.4& 0.97\\
       0 & 0       &325.3& 11.6& 905.8&29.3&515.3&14.1\\
       0 & 0       &8.3&2.3&29.3&5.4&16.1&2.9\\
       0& 0        &157.4& 6.4&515.2&16.1& 1375.1& 40.4\\
       0&0         &3.6&0.97& 14.1& 2.9& 40.4& 6.3\\
      \end{array}
    \right)$ is the covariance matrix, and
\begin{equation}\label{}
\begin{split}
  s_{BAO}=&(\frac{r_{\rm d}}{D_{V}(0.106)}-0.336,\frac{r_{\rm d}}{D_{V}(0.15)}-0.2239,D_{M}(0.38)(147.78/r_{\rm d})-1512.39,3\times10^5\times H(0.38)(r_{\rm d}/147.78)\\&-81.2087, D_{M}(0.51)(147.78/r_{\rm d})-1975.22,3\times10^5\times H(0.51)(r_{\rm d}/147.78)-90.9029,D_{M}(0.61)(147.78/r_{\rm d})\\&-2306.68,3\times10^5\times H(0.61)(r_{\rm d}/147.78)-98.9647).
\end{split}
\end{equation}
\end{widetext}
where $D_{\rm V}(z)$ is the dilation scale and is given by
\begin{equation}\label{}
  D_{\rm V}(z)=(z D_{\rm M}(z)^{2}/H(z))^{\frac{1}{3}},
\end{equation}
$r_{\rm d}=r_{\rm s}(z_{\rm d})$ is the sound horizon at the drag epoch $z_{\rm d}$, $z_{\rm d}$ can be calculated by using \cite{Eisenstein1997Baryonic}
\begin{eqnarray}
  z_{\rm d} &=&1291\frac{\omega_{\rm m}^{0.251}}{1+0.659\omega_{\rm m}^{0.828}}(1+b_{1}\omega_{\rm b}^{b_{2}}) \\
  b_1 &=& 0.313\omega_{\rm m}^{-0.419}(1+0.607\omega_{\rm m}^{0.674})\\
  b_2&=&0.238\omega_{\rm m}^{0.223}
\end{eqnarray}

\subsection{Type Ia supernovae}
For SN Ia observations, we consider the following likelihood that consists of 6 effective points on the Hubble parameter $E(z)$.
\begin{equation}
 \chi_{\rm SN}^{2}=s_{\rm SN}^{T} C_{\rm SN}^{-1}s_{\rm SN},
\end{equation}
\begin{equation}
\begin{split}
  s_{\rm SN}= & (\frac{1}{E(0.07)}-1.003,\frac{1}{E(0.2)}-0.901,\frac{1}{E(0.35)}-0.887,\\&\frac{1}{E(0.55)}-0.732,
     \frac{1}{E(0.9)}-0.656,\frac{1}{E(1.5)}-0.342,)
\end{split}
\end{equation}
where $C_{ij}=D_{ij}\sigma_i\sigma_j$ is the covariance matrix, $\sigma_{\rm SN}=(0.023,0.017,0.029,0.033,0.052,0.079)$ is the errors,
and
$D_{\rm SN}=  \left(
      \begin{array}{cccccc}
        1 & 0.39 & 0.53&0.37&0.01&-0.02\\
       0.39 & 1 & -0.14&0.37&-0.08&-0.08\\
       0.53& -0.14&  1 &-0.16&0.17&-0.07\\
       0.37 &0.37&-0.16&  1  &-0.39&0.15\\
       0.01 & -0.08&0.17&-0.39 & 1 &-0.19\\
       -0.02&-0.08&-0.07& 0.15&-0.19 & 1\\
      \end{array}
    \right)$
is the covariance.

Such likelihood compress most of the information of 1048 SN Ia contained in the Pantheon compilation \cite{scolnic2018complete} and the 15 SN Ia at $z > 1$ from the Hubble Space Telescope Multi-Cycle Treasury programs \cite{riess2018type}.

\subsection{Results}
\begin{figure*}
	\centering
	\includegraphics[scale=0.35]{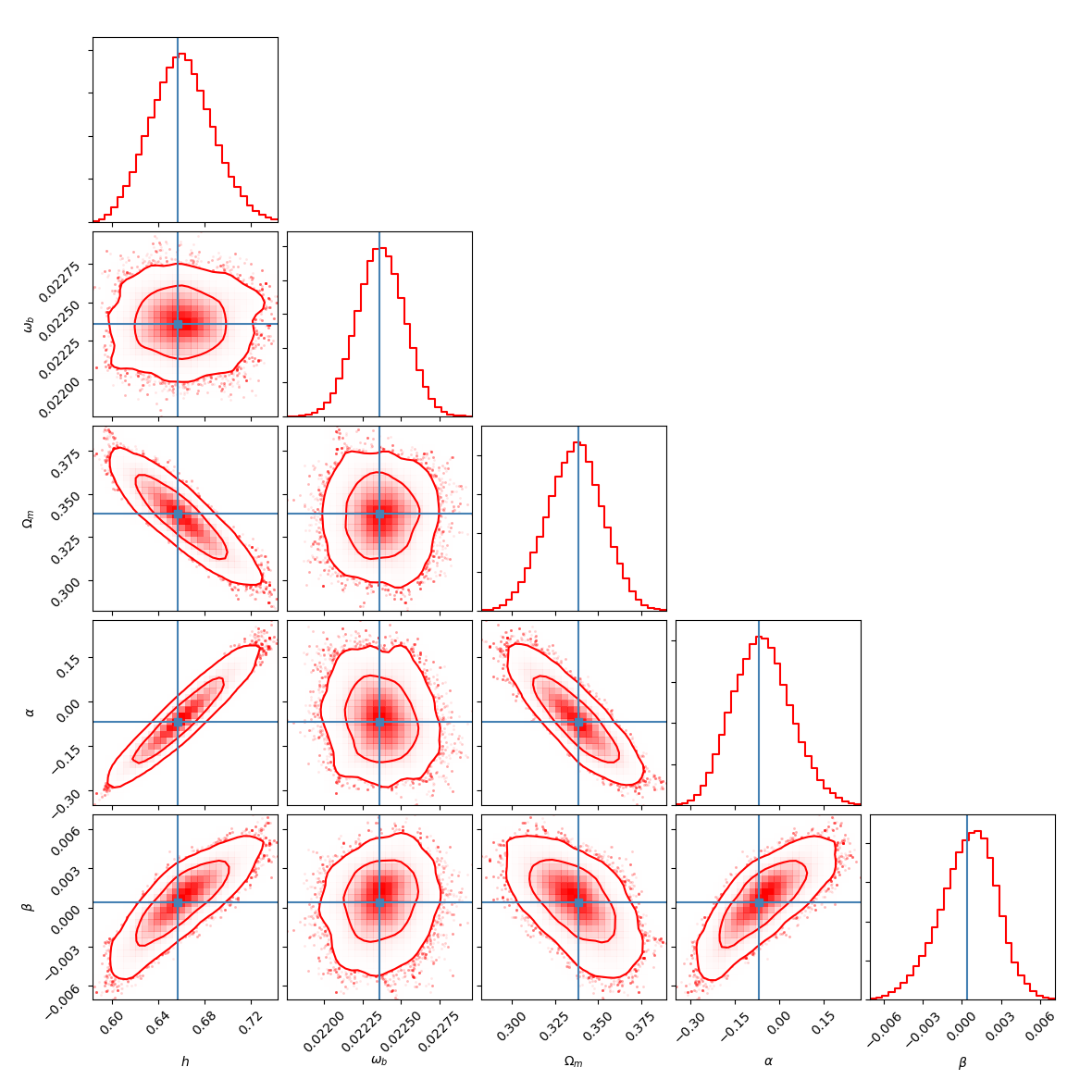}
	\caption{one dimensional posterior distribution and two dimensional joint contours for the parameters in the parametric interacting dark energy model specified by $k=0$ using CMB+BAO data set.}\label{fig:3}
\end{figure*}

\begin{figure*}
	\centering
	\includegraphics[scale=0.35]{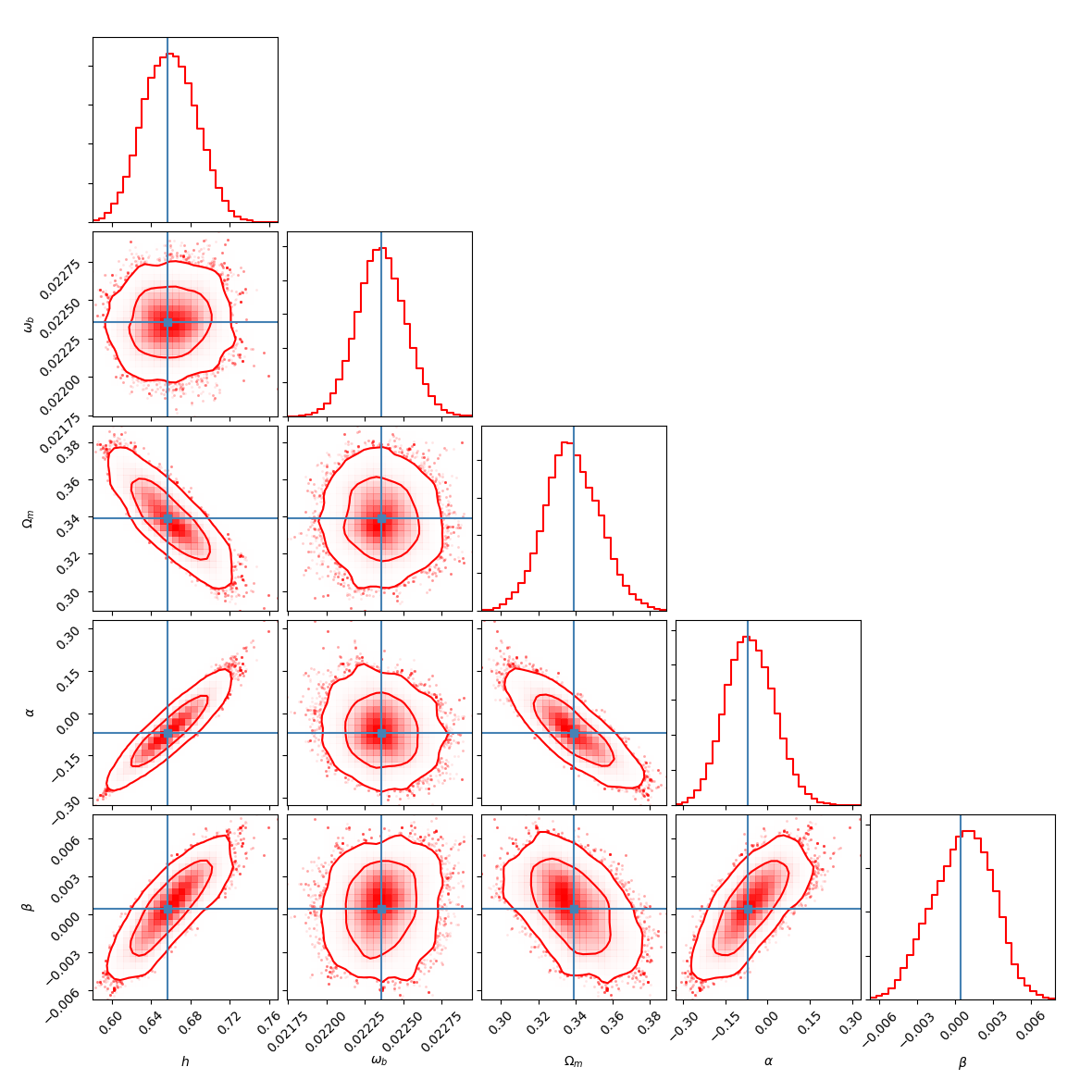}
	\caption{one dimensional posterior distribution and two dimensional joint contours for the parameters in the parametric interacting dark energy model specified by $k=5$ using CMB+BAO data set.}\label{fig:4}
\end{figure*}

\begin{figure*}
	\centering
	\includegraphics[scale=0.35]{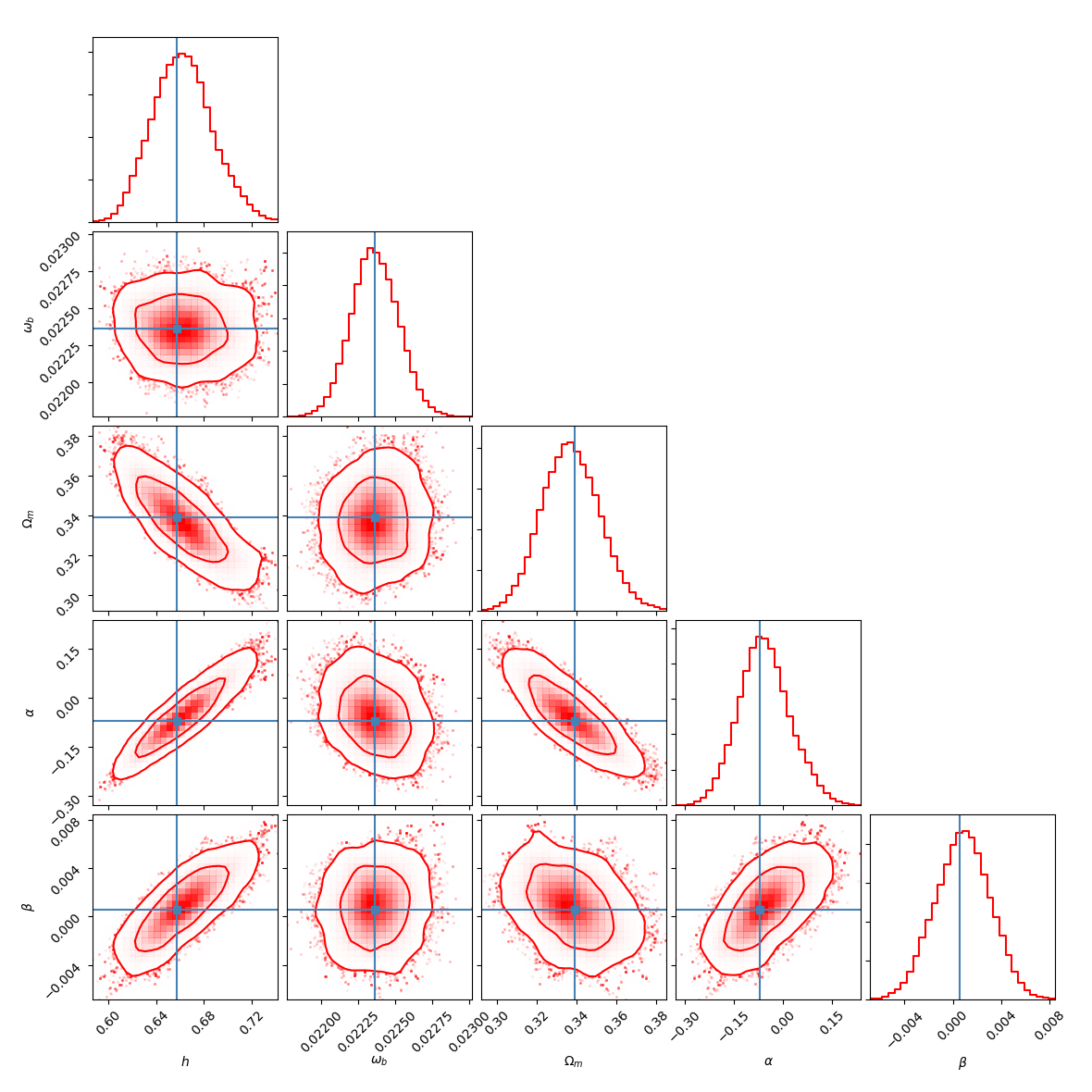}
	\caption{one dimensional posterior distribution and two dimensional joint contours for the parameters in the parametric interacting dark energy model specified by $k=10$ using CMB+BAO data set.}\label{fig:5}
\end{figure*}

\begin{figure*}
	\centering
	\includegraphics[scale=0.35]{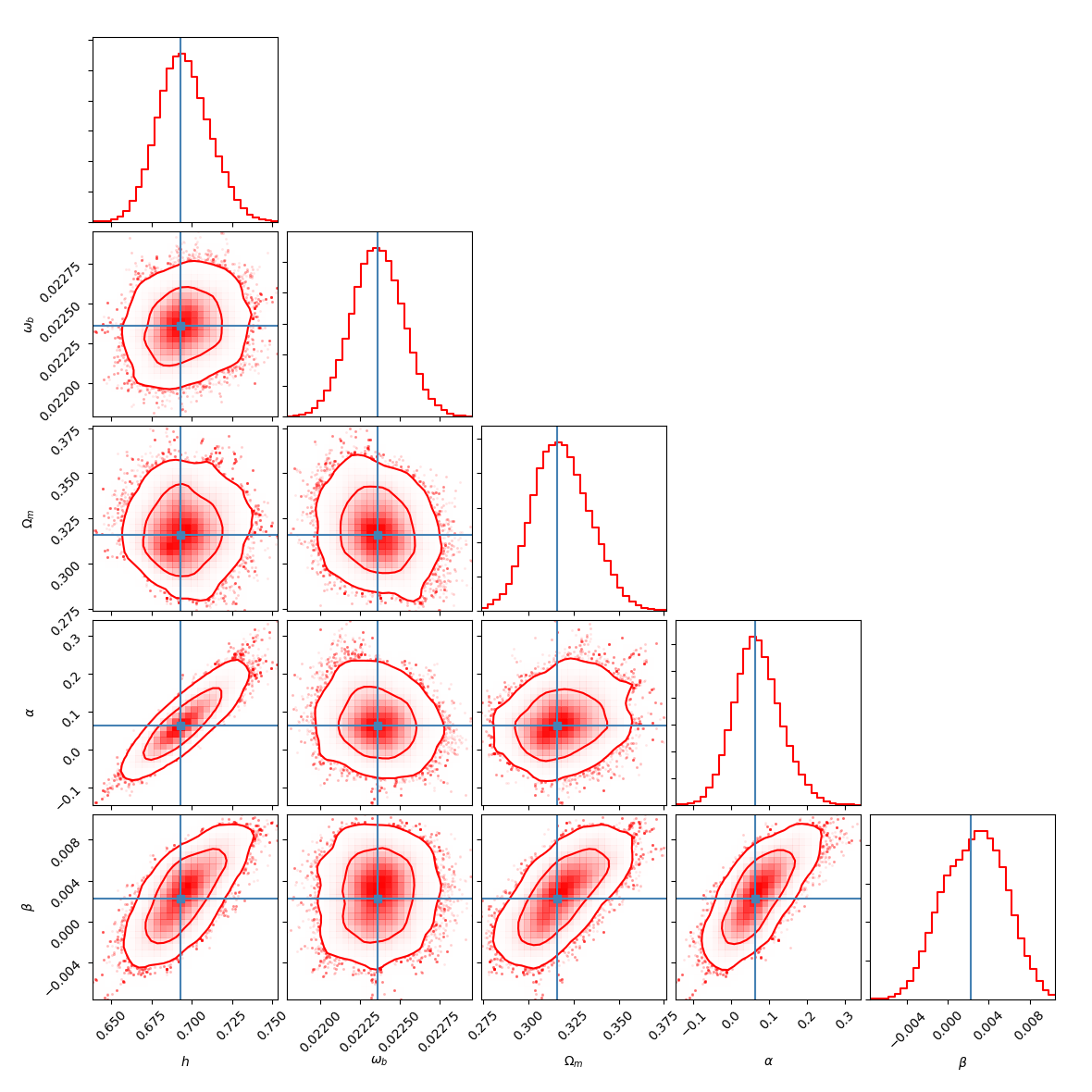}
	\caption{one dimensional posterior distribution and two dimensional joint contours for the parameters in the parametric interacting dark energy model specified by $k=0$ using CMB+Pantheon data set.}\label{fig:6}
\end{figure*}

\begin{figure*}
	\centering
	\includegraphics[scale=0.35]{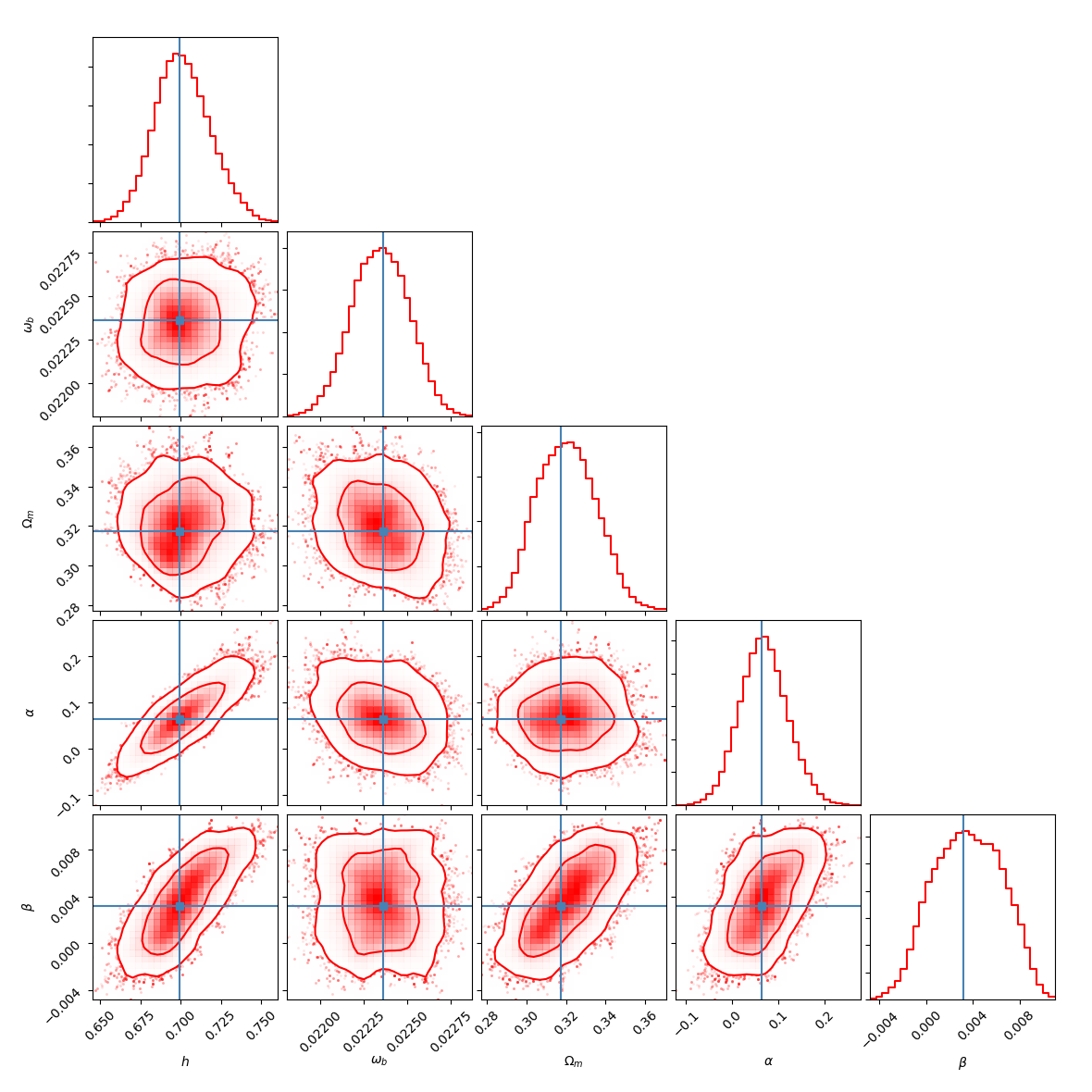}
	\caption{one dimensional posterior distribution and two dimensional joint contours for the parameters in the parametric interacting dark energy model specified by $k=5$ using CMB+Pantheon data set.}\label{fig:7}
\end{figure*}

\begin{figure*}
	\centering
	\includegraphics[scale=0.35]{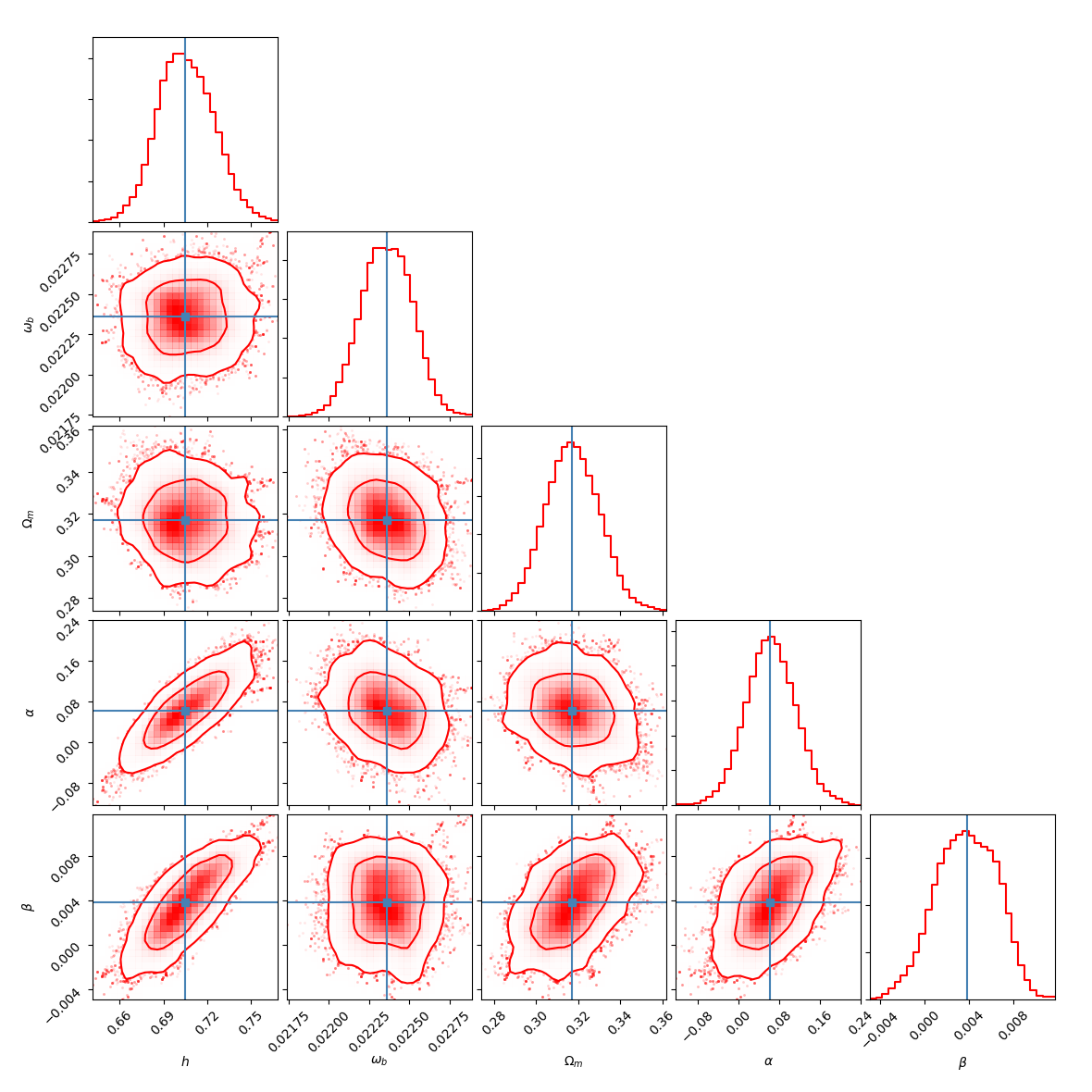}
	\caption{one dimensional posterior distribution and two dimensional joint contours for the parameters in the parametric interacting dark energy model specified by $k=10$ using CMB+Pantheon data set.}\label{fig:8}
\end{figure*}

\begin{figure*}
	\centering
	\includegraphics[scale=0.35]{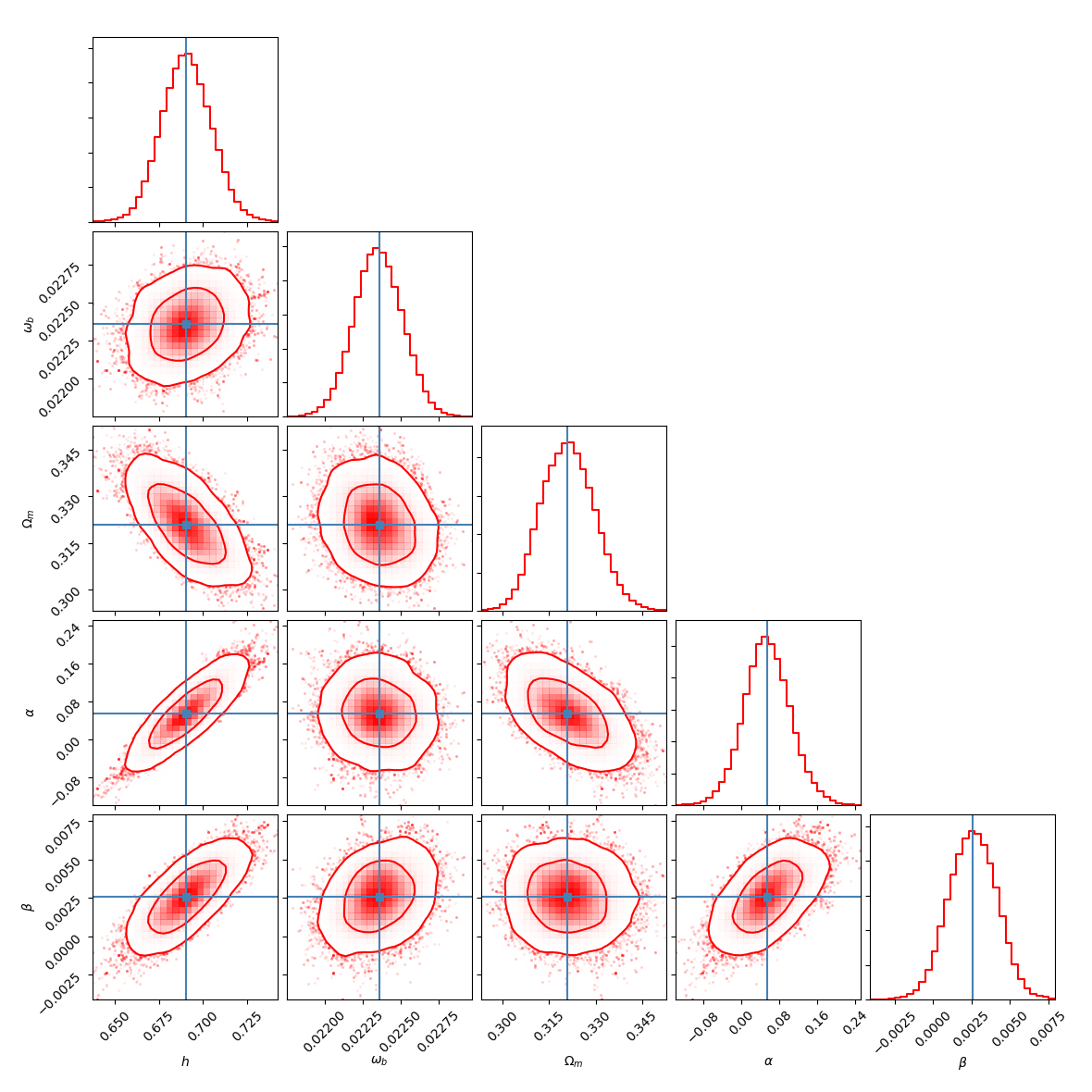}
	\caption{one dimensional posterior distribution and two dimensional joint contours for the parameters in the parametric interacting dark energy model specified by $k=0$ using CMB+BAO+Pantheon data set.}\label{fig:9}
\end{figure*}

\begin{figure*}
	\centering
	\includegraphics[scale=0.35]{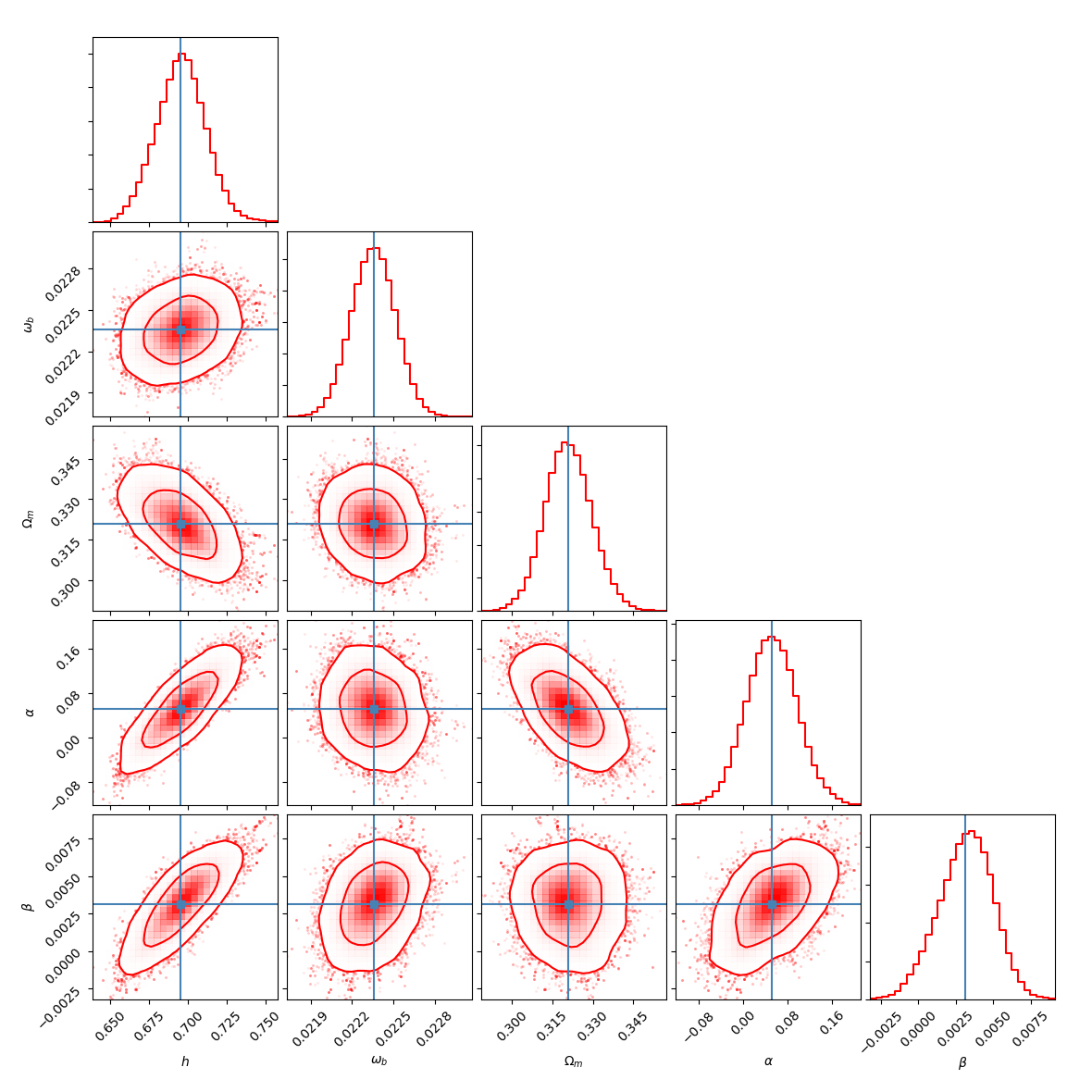}
	\caption{one dimensional posterior distribution and two dimensional joint contours for the parameters in the parametric interacting dark energy model specified by $k=5$ using CMB+BAO+Pantheon data set.}\label{fig:10}
\end{figure*}

\begin{figure*}
	\centering
	\includegraphics[scale=0.35]{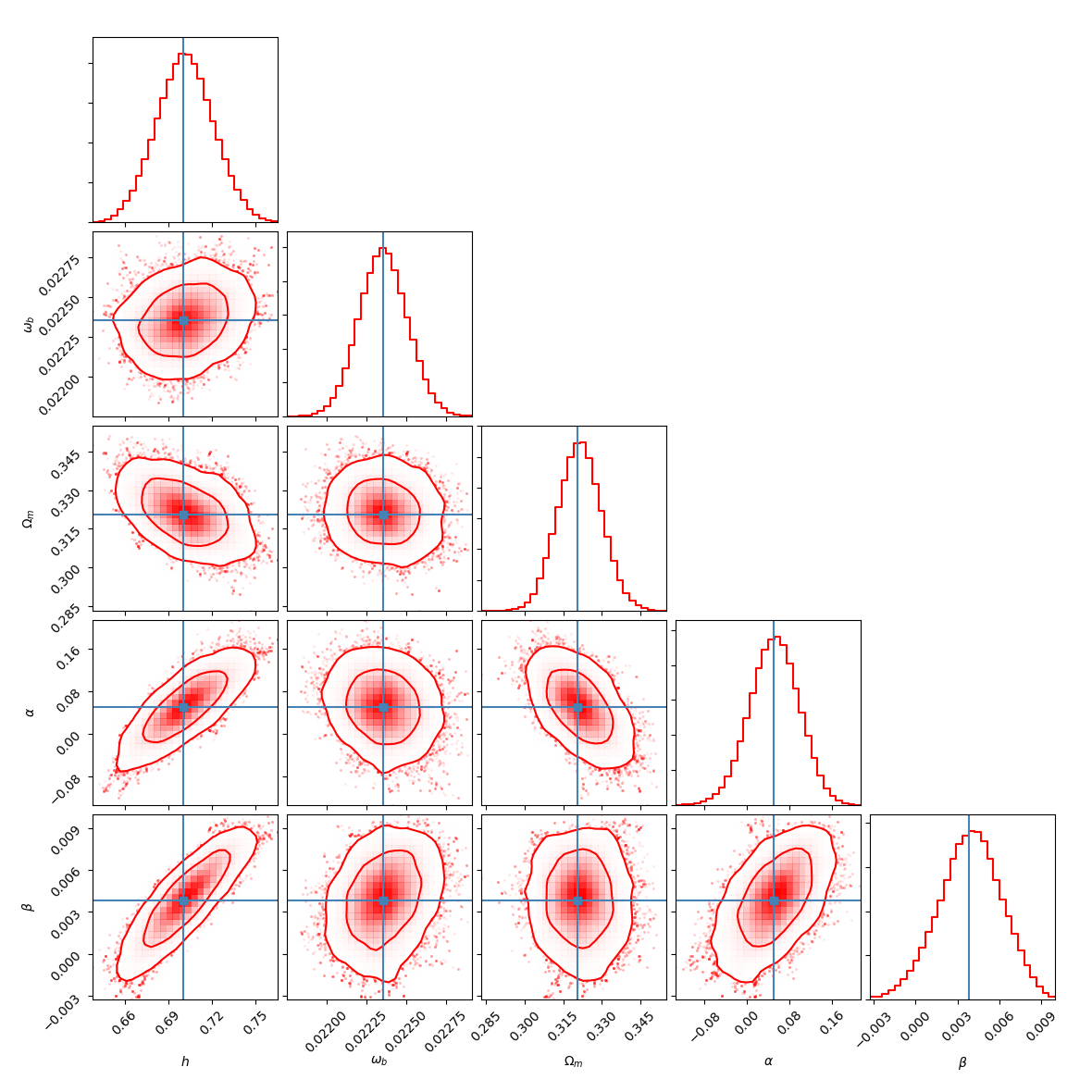}
	\caption{one dimensional posterior distribution and two dimensional joint contours for the parameters in the parametric interacting dark energy model specified by $k=10$ using CMB+BAO+Pantheon data set.}\label{fig:11}
\end{figure*}

In Tab.\ref{tab:1}-\ref{tab:3} and Fig.\ref{fig:3}-\ref{fig:11}, we summarize the constraints on three parametric interacting dark energy model specified by $k=0,5,10$ for various cosmological probes, namely, CMB+BAO, CMB+Pantheon, CMB+BAO+Pantheon. One notes that using CMB distance priors alone to fit the models are not under our consideration since the number of data points in the CMB distance priors data sets are less than the number of free parameters in each model.

For the CMB+BAO data set, the fitting results of $\alpha$ for three parametric models specified by $k=0,5,10$ are $\alpha=-0.059_{-0.095}^{+0.099},-0.064_{-0.081}^{+0.086},-0.059_{-0.072}^{+0.079}$ at 68\% CL respectively, all in agreement with $\alpha=0$ within 1$\sigma$. And the fitting results of $\beta$ for three parametric models specified by $k=0,5,10$ are $\beta=0.0006_{-0.0023}^{+0.0019},0.0007_{-0.0026}^{+0.0023},0.0008_{-0.0023}^{+0.0022}$ at 68\% CL respectively, also all in agreement with $\beta=0$ within 1$\sigma$. Finally, the fitting results of $H_0$ for three parametric models specified by $k=0,5,10$ are $H_0=0.660_{-0.026}^{+0.027},0.659_{-0.026}^{+0.027},0.662_{-0.024}^{+0.025}$ at 68\% CL, which relieve the Hubble tension to 2.4$\sigma$, 2.5$\sigma$ and 2.5$\sigma$ respectively. The results are as we expected, since as pointed out by Di Valentino et. al. in \cite{di2021realm}, the solution to the Hubble tension can introduce a further disagreement with the BAO data, which strongly weaken the ability to relieve the Hubble tension for a late-time scenario by using CMB+BAO data set.

For the CMB+Pantheon data set, the fitting results of $\alpha$ for three parametric models specified by $k=0,5,10$ are $\alpha=0.068_{-0.056}^{+0.063},0.068_{-0.048}^{+0.051},0.064_{-0.047}^{+0.048}$ at 68\% CL respectively, indicating a positive $\alpha$ with a significance above 1$\sigma$ for each model, considering $w_{\rm de}=-1-\alpha$, one finds that CMB+Pantheon data set supports a phantom scenario with a significance above 1$\sigma$. And the fitting results of $\beta$ for three parametric models specified by $k=0,5,10$ are $\beta=0.003_{-0.003}^{+0.003},0.004_{-0.003}^{+0.003},0.004_{-0.003}^{+0.003}$ at 68\% CL respectively, showing a non-zero effective dark matter equation of state with a significance above 1$\sigma$ for each model. The results is reasonable, since the nature of phantom dark energy decrease their density with respect to redshift $z$, considering parameter $\alpha$ is uncorrelated with the dark matter density parameter $\Omega_m$ for the CMB+Pantheon data set, which can be inferred from Fig.\ref{fig:6}-\ref{fig:8}, faster growth trends with respect to redshift $z$ for dark matter density is necessary for each model to be compatible with the low-redshift observation, i.e. the Pantheon data set, which suggests a "more positive" $\beta$ compared to the model in which dark energy is played by a cosmological constant. Since $\alpha$ and $\beta$ both shift to "more positive" values for this data set, we obtain three larger values of Hubble constant,
namely $H_0=0.694_{-0.015}^{+0.017},0.701_{-0.016}^{+0.018},0.705_{-0.017}^{+0.020}$ at 68\% CL, which relieve the Hubble tension to 1.8$\sigma$, 1.4$\sigma$ and 1.1$\sigma$,
revealing that $k$ is positive correlated to the Hubble constant for the CMB+Pantheon data set.

For the CMB+BAO+Pantheon data set, the best fitting values of parameters $\alpha$, $\beta$ lie between their counterparts inferred from CMB+BAO data set and CMB+Pantheon data set. Just like the CMB+Pantheon case, the full data set also prefer a positive $\alpha$ and a non-zero effective dark matter equation of state both with a significance above 1$\sigma$ for each model. The fitting results of $H_0$ for three parametric models specified by $k=0,5,10$ are $H_0=0.691_{-0.014}^{+0.014},0.696_{-0.016}^{+0.015},0.702_{-0.019}^{+0.019}$ at 68\% CL, which relieve the Hubble tension to 2.3$\sigma$, 1.9$\sigma$ and 1.3$\sigma$ respectively, revealing that $k$ is positive correlated to the Hubble constant for the full data set. Unlike the CMB+BAO case, adding SN Ia data set shifts the $\beta$ to larger values, which modifies early-time evolution history of the universe, therefore raising the Hubble constant introduces less disagreement with the BAO data in this case.
Finally, we obtain $\chi_{\rm min}^2=10.43,10.47,10.48$ for three parametric models specified by $k=0,5,10$ respectively, showing no model prefers over the rest.
\begin{table*}[!t]
	\renewcommand\arraystretch{1.5}
	\caption{68\% CL constraints on the parametric interacting dark energy model specified by $k=0$ in light of CMB, BAO and Pantheon data sets.}
	\setlength{\tabcolsep}{3mm}{
		{\begin{tabular}{c c c c}
               \hline
	            \hline
 Parameters & ${\rm CMB+BAO}$ & ${\rm CMB+Pantheon}$&${\rm CMB+BAO+Pantheon}$
\\ \hline
$\Omega_{\rm m}$    &$0.337_{-0.016}^{+0.016}$ &  $ 0.318_{-0.014}^{+0.017}$ & $ 0.321_{-0.008}^{+0.009}$
                     \\
$\omega_{\rm b}$   &    $0.02237_{-0.00015}^{+0.00015}$ &  $0.02236_{-0.00016}^{+0.00016}$& $ 0.02235_{-0.00015}^{+0.00016}$
                       \\
 $h$          &    $ 0.660_{-0.026}^{+0.027}$ & $0.694_{-0.015}^{+0.017}$& $ 0.691_{-0.014}^{+0.014}$
                       \\
$\alpha$         &  $-0.059_{-0.095}^{+0.099}$ &  $ 0.068_{-0.056}^{+0.063}$& $0.054_{-0.045}^{+0.047} $
                     \\
$\beta$         &  $0.0006_{-0.0023}^{+0.0019}$ & $0.003_{-0.003}^{+0.003}$& $ 0.003_{-0.002}^{+0.002}$
                     \\
				\hline
				\hline
		    \end{tabular}
			\label{tab:1}}}
\end{table*}

\begin{table*}[!t]
	\renewcommand\arraystretch{1.5}
	\caption{68\% CL constraints on the parametric interacting dark energy model specified by $k=5$ in light of CMB, BAO and Pantheon data sets.}
	\setlength{\tabcolsep}{3mm}{
		{\begin{tabular}{c c c c}
               \hline
	            \hline
 Parameters & ${\rm CMB+BAO}$ & ${\rm CMB+Pantheon}$&${\rm CMB+BAO+Pantheon}$
\\ \hline
$\Omega_{\rm m}$    &$0.338_{-0.013}^{+0.016}$ &  $ 0.317_{-0.013}^{+0.015}$ & $ 0.321_{-0.008}^{+0.009}$
                     \\
$\omega_{\rm b}$   &    $0.02236_{-0.00015}^{+0.00016}$ &  $0.02234_{-0.00016}^{+0.00016}$& $ 0.02235_{-0.00015}^{+0.00015}$
                       \\
 $h$          &    $ 0.659_{-0.026}^{+0.027}$ & $0.701_{-0.016}^{+0.018}$& $ 0.696_{-0.016}^{+0.015}$
                       \\
$\alpha$         &  $-0.064_{-0.081}^{+0.086}$ &  $ 0.068_{-0.048}^{+0.051}$&  $0.051_{-0.044}^{+0.044}$
                     \\
$\beta$         &  $0.0007_{-0.0026}^{+0.0023}$ & $0.004_{-0.003}^{+0.003}$& $ 0.003_{-0.002}^{+0.002}$
                     \\
				\hline
				\hline
		    \end{tabular}
			\label{tab:2}}}
\end{table*}

\begin{table*}[!t]
	\renewcommand\arraystretch{1.5}
	\caption{68\% CL constraints on the parametric interacting dark energy model specified by $k=10$ in light of CMB, BAO and Pantheon data sets.}
	\setlength{\tabcolsep}{3mm}{
		{\begin{tabular}{c c c c}
               \hline
	            \hline
 Parameters & ${\rm CMB+BAO}$ & ${\rm CMB+Pantheon}$&${\rm CMB+BAO+Pantheon}$
\\ \hline
$\Omega_{\rm m}$    &$0.337_{-0.014}^{+0.014}$ &  $ 0.317_{-0.012}^{+0.013}$ & $ 0.321_{-0.008}^{+0.008}$
                     \\
$\omega_{\rm b}$   &    $0.02236_{-0.00014}^{+0.00016}$ &  $0.02236_{-0.00015}^{+0.00015}$& $ 0.02236_{-0.00014}^{+0.00015}$
                       \\
 $h$          &    $ 0.662_{-0.024}^{+0.025}$ & $0.705_{-0.017}^{+0.020}$& $ 0.702_{-0.019}^{+0.019}$
                       \\
$\alpha$         &  $-0.059_{-0.072}^{+0.079}$ &  $ 0.064_{-0.047}^{+0.048}$&  $0.052_{-0.046}^{+0.044}$
                     \\
$\beta$         &  $0.0008_{-0.0023}^{+0.0022}$ & $0.004_{-0.003}^{+0.003}$& $ 0.004_{-0.002}^{+0.002}$
                     \\
				\hline
				\hline
		    \end{tabular}
			\label{tab:3}}}
\end{table*}

\section{Conclusions}
With a difference of about 5$\sigma$ level between $H_0$ restricted from Planck 2018 CMB anisotropy data assuming the $\Lambda$CDM model and the latest distance measurements of SN Ia calibrated by Cepheid variables, i.e. R22, the Hubble constant problem now is increasingly being taken seriously by the cosmology community. In order to relieve the Hubble tension, in this paper we propose a new interacting dark energy model by parameterizing the densities of dark matter and dark energy, which can be consider as a interacting dark energy model with time varying coupling parameter, this parametric approach for interacting dark sectors are inspired by our previous work concerning the coupled generalized three-form dark energy model in which dark matter and dark energy behave like two uncoupled dark sectors with effective equation of state when the three-form $|\kappa X|\gg1$. For the purpose to connect such parametric model with coupled generalized three-form dark energy, we reconstruct coupled generalized three-form dark energy from such parametric model under the condition $|\kappa X_0|\gg1$, by computing the evolution history of the average effective equation of state of dark energy and dark matter, we find that the parametric model can be regarded as the generalization of the coupled generalized three-form dark energy we proposed in \cite{yao2021relieve}. We then place constraints on three parametric models specified by $k=0,5,10$ in light of CMB distance priors, BAO measurements and Pantheon, and find that for the CMB+BAO data set, the fitting results of $H_0$ for these three models are all relieve the Hubble tension to $\sim2.5\sigma$, showing no correlation between $k$ and $H_0$ for this data set. And for the CMB+Pantheon and CMB+BAO+Pantheon data sets, fitting results prefer a positive $\alpha$ and a non-zero effective dark matter equation of state both with a significance above 1$\sigma$ for each model. In addition, there are some evidences supporting a positive correlation between $k$ and $H_0$, for instance, providing the full data set, the fitting results of $H_0$ for three parametric models specified by $k=0,5,10$ are $H_0=0.691_{-0.014}^{+0.014},0.696_{-0.016}^{+0.015},0.702_{-0.019}^{+0.019}$ at 68\% CL, which relieve the Hubble tension to 2.3$\sigma$, 1.9$\sigma$ and 1.3$\sigma$ respectively, indicating that a non-zero $k$ is beneficial for relieving the Hubble tension. Finally, considering $\chi_{\rm min}^2=10.43,10.47,10.48$ obtained by using the full data set for three parametric models specified by $k=0,5,10$, no model is preferred over by another one.
\section*{Acknowledgments}
The paper is partially supported by the Natural Science Foundation of China.

\bibliographystyle{spphys}
\bibliography{mgy}

\end{document}